\documentclass[useAMS,usenatbib]{mn2e}
\usepackage{psfig}

\newcommand\msun{{\,M_\odot}}

\newcommand\fe{Fe K$\alpha$\ }

\newcommand\sgra{Sgr~A$^*$}

\def\>{$>$}
\def\<{$<$}

\newcommand\simlt{\lower.5ex\hbox{$\; \buildrel < \over \sim \;$}}
\newcommand\simgt{\lower.5ex\hbox{$\; \buildrel > \over \sim \;$}}
\def\sqr#1#2{{\vcenter{\hrule height.#2pt
      \hbox{\vrule width.#2pt height#1pt \kern#1pt
         \vrule width.#2pt}
      \hrule height.#2pt}}}

\def\del#1{{}}

\title[]{Warped accretion disks and the unification of \\
Active Galactic Nuclei}

\author[Nayakshin]{Sergei Nayakshin \\ Max-Planck-Institut f\"ur
Astrophysik, Karl-Schwarzschild-Str. 1, 85740 Garching, Germany}
\date{2004 Xxxxx XX}

\pagerange{\pageref{firstpage}--\pageref{lastpage}} \pubyear{2003}

\def\LaTeX{L\kern-.36em\raise.3ex\hbox{a}\kern-.15em
    T\kern-.1667em\lower.7ex\hbox{E}\kern-.125emX}

\begin{document}

\label{firstpage}

\maketitle

\begin{abstract}
Orientation of parsec-scale accretion disks in AGN is likely to be
nearly random for different black hole feeding episodes. Since AGN
accretion disks are unstable to self-gravity on parsec scales, star
formation in these disks will create young stellar disks, similar to
those recently discovered in our Galactic Center. The disks blend into
the quasi-spherical star cluster enveloping the AGN on time scales
much longer than a likely AGN lifetime. Therefore, the gravitational
potential within the radius of the black hole influence is at best
axi-symmetric rather than spherically symmetric. Here we show that as a
result, a newly formed accretion disk will be warped. For the simplest
case of a potential resulting from a thin stellar ring, we calculate
the disk precession rates, and the time dependent shape. We find that,
for a realistic parameter range, the disk becomes strongly warped in
few hundred orbital times.  We suggest that this, and possibly other
mechanisms of accretion disk warping, have a direct relevance to the
problem of AGN obscuration, masing warped accretion disks, narrow \fe
lines, etc.
\end{abstract}

\section{Introduction}

Observations show that a significant fraction, perhaps a majority, of
AGN of different types are obscured by a screen of a cold dusty
matter, thought to be a molecular torus-like structure with a scale
between 1 to a 100 parsec
\citep{Antonucci85,Antonucci93,Maiolino95,Maiolino99,Sazonov04,Jaffe04}.
Moreover, many lines of observational evidence suggest that the
unobscured AGN have similar torii as well, but we are viewing these
AGN at an angle allowing a direct view of their central engines. The
orientation-dependent obscuration is thus said to unify the different
AGN classes \citep{Antonucci85,Antonucci93}.

Despite the observational importance, the theory of molecular torii is
still in an exploratory state, which is indicative of the difficulty
of the problem.  \cite{Krolik88} \citep[see also][]{Krolik86} showed
that the large geometrical thickness of the torus and yet its
apparently low temperature are consistent only if the torus is made of
many molecular clouds moving with very high random velocities (Mach
number $\simgt 30$). Given that stellar feedback processes appear
inefficient to explain these large random speeds, the authors were
``forced to seek a much more speculative solution: viscous heating of
the cloud system due to partially elastic cloud-cloud collisions.''
However, until now there is no definitive answer (via numerical
simulations) on whether clouds colliding at large Mach numbers would
behave even partially elastic, or would rather share and dump the
random velocity component and collapse to a disk configuration.

\cite{Wada02} took a point with \cite{Krolik88} assertion of the
weakness of the stellar processes. Considering ``large'', e.g. 100
parsec torii, and via numerical simulations of supernova explosions
resulting from star formation in a self-gravitating, massive disk,
they have shown that large random cold gas velocities result from the
interaction of the gas with the supernova shells.  However as such
about $99$~\% of the gaseous disk in the simulations is consumed in
the star formation episode rather than being accreted by the black
hole. Unless most of the stars formed in the disk are later somehow
accreted onto the SMBH, this mechanism can only work for torii on
scales much larger than the SMBH gravitational sphere of influence,
$R_h$, i.e. where the stellar mass is much greater than $M_{\rm
BH}$. It is also not clear whether this mechanism would work for
smaller accretion/star formation rates because then supernova
explosions become too rare. Finally, time variability of obscuring
column depths 
argues for much smaller radial sizes of the ``torus'', e.g. in the
range $0.1-10$ parsec \citep{Risaliti02}.

Here we would like to emphasize that accretion disks in AGN are very
unlikely to be planar. There are several mechanisms that are capable
of producing strong warps. We believe that such warps have to be an
integral part of the AGN obscuration puzzle. In particular, we shall
describe and calculate a disk warping mechanism driven by an
axi-symmetric gravitational potential.

The sequence of events that takes place in an AGN feeding cycle in our
model is as following. First of all, a merger with another galaxy or a
satellite, or another source of cold gas, fills the inner part of the
galaxy with plenty of gas. This gas will in general have a significant
angular momentum oriented practically {\em randomly}.  The gas will
start settling in a disk that is too massive and too cold to be stable
against self-gravity
\citep[e.g.,][]{Paczynski78,Shlosman89,Collin99,Goodman03}. Stars are formed
inside the accretion disk, producing a flat stellar system. This
long-lived axi-symmetric (or perhaps also warped) structure will
torque any orbit which is not exactly co-planar with it, resulting in
a precession around the symmetry axis. Now, as the time goes on, the
orientation of the angular momentum vector of the incoming gas
changes, and the newly built disk is exposed to the torque from the
stellar disk remnant. Different rings of the disk precess at different
rates, thus the disk becomes warped.

We shall emphasize that existence of these stellar disk remnants is
hardly a question based on the severe self-gravity problems for the
standard accretion disks at large radii. The recent observational
evidence supports this conclusion too. The best known example of such
flat stellar systems is in our Galactic Center \citep{Genzel03} where
two young stellar rings are orbiting the SMBH only a tenth of a parsec
away. The orbital planes of these rings are oriented at very large
angles to the Galactic plane. Such flat stellar systems are also being
found elsewhere \citep[e.g.,][]{KJ04}.

In this paper, we concentrate on the linear gravitational warping
effect to investigate its main features. Taking the simplest case of a
warp produced by a massive ring, we calculate the gravitational
warping torque. Starting from a thin test accretion disk, we calculate
its time dependent shape. Under quite realistic assumptions, the disk
becomes strongly warped in some $10^2-10^3$ orbital times, i.e. in
$10^5-10^7$ years. Since the warping is gravitational in nature,
gaseous and clumpy molecular disks alike are subject to such
deformations. We speculate that in the non-linear stage of the effect,
a clumpy warped disk will form a torus-like structure.

\section{Torques in a linear regime}


The basic physics of the effect is very simple and has been explained
by e.g., \cite{Binney92} in his consideration of the galactic warps
(most galactic disks are somewhat warped). Consider a nearly circular
orbit of radius $R$ for a test particle in an axi-symmetric potential
(e.g. a central point mass plus a flat disk). Let the axis $z$ to be
perpendicular to the plane of the disk. The particle makes radial and
vertical oscillations with slightly different frequencies, and the
difference is the precession frequency, $\omega_p$.  Due to the
symmetry, the $z$-projection of the angular momentum of the particle
is exactly conserved \citep[see, e.g., \S 3.2 in][]{BT87}.  Thus the
angle between the angular momentum of the particle and the $z$-axis,
$\beta$, is a constant. The line of the nodes (the line over which the
two planes intersect), however, precesses. 

Now consider a ``test-particle'' disk initially in the same plane as
the circular orbit. The disk is a collection of rings, i.e. circular
orbits. Since precession rates $\omega_p$ are different for different
$R$, rings turn around the $z$-axis on unequal angles. The initially
flat disk will be warped with time.

\subsection{Gravitational torques between two rings}\label{sec:rings}

Let us calculate the gravitational torque between two rings with radii
$R_1$ and $R$, inclined at angle $\beta$ with respect to each
other. We work in two rigid coordinate systems, $x,y,z$ and
$x',y',z'$, centers of which are at the SMBH.  For convenience, we
place the first ring in the $z=0$ plane, whereas the second ring is at
$z'=0$ plane. Angle $\beta$ is obviously the angle between the axes $z$
and $z'$. Further, the axes $x$ and $x'$ are chosen to coincide with
the line of the nodes.

The total torque exerted by the second ring on the first one is
\begin{equation}
\vec{\tau}_{21} = G \sigma_1 R_1 \sigma R \int_0^{2\pi} d\phi_1 \;
\int_0^{2\pi} d\phi \frac{\; [\vec{r}_1 \times
\vec{r}]}{|\vec{r} - \vec{r}_1|^{3}}
\end{equation}
where $\sigma_1=M_1/2\pi R_1$ and $\sigma = M/2\pi R$, with $M_1$ and
$M$ being the masses of the two rings, respectively. The integration
goes over angles $\phi_1$ and $\phi$ that are azimuthal angles in the
respective frames of the rings.  From this expression it is
immediately clear that co-planar rings do not exert any torque on each
other, $[\vec{r}_1\times \vec{r}]=0$. In addition, if $\beta =
\pi/2$, the integral vanishes as well because for each $\vec{r}_1$ the
opposites sides of the second ring (e.g. $\phi$ and $\phi+ \pi$) make
equal but opposing contributions.

It is also possible to show that due to symmetry the torque's
$z$-projection vanish. Thus we only have the $\tau_{21,x} =
-\tau_{12,x}$ component, meaning that the angular momentum vector of
the ring will rotate without changing its magnitude. Also, if $M_1 \gg
M$, then one can neglect warping of the first ring.

The absolute distance between two ring elements with the respective
angles $\phi_1$ and $\phi$ is
\begin{equation}
\left[\vec{r}_1 - \vec{r}\right]^2 = R_1^2 + R^2 -2 R_1 R
\cos\lambda\;,
\end{equation}
where 
\begin{equation}
\cos\lambda \; = \; \cos\beta \sin\phi \sin\phi_1 +
\cos\phi_1\cos\phi\;.
\end{equation}
Without going into tedious detail, we write the torque expression
separating out the leading radial dependence and the integral over
angles, which we label $I(\delta, \beta)$:
\begin{equation}
\tau_{21,x} = \frac{G M_1 M R_1 R}{ (R_1^2 + R^2)^{3/2}} \; I(\delta,
\beta)\;,
\end{equation}
with
\begin{equation}
I(\delta, \beta) \; \equiv \sin\beta \int_0^{2\pi} \frac{d\phi_1}{2\pi}
\int_0^{2\pi} \frac{d\phi}{2\pi}\; \frac{\sin\phi_1\sin\phi}{\left[1
- \delta\cos\lambda\right]^{3/2}}
\label{i}
\end{equation}
and
\begin{equation}
\delta \equiv \frac{2 R_1 R}{R_1^2 + R^2}\;.
\end{equation}

Total angular momentum of the second ring is $L= M \Omega_K R^2$.
Recall that $L_z = L \cos\beta =$~const, whereas the component of
$\vec{L}$ in the plane of the first ring ($z=0$) precesses. We chose
it to be initially in the $y$-direction, so that $L_y(t=0) = L
\sin\beta$ (cf. equation \ref{defl} below). The precession frequency
for the second ring is then
\begin{equation}
\omega_p = \frac{\tau_{12,x}}{L_y} = \frac{\tau_{12,x}}{M \Omega_K R^2
\sin\beta}\;,
\label{omegaex}
\end{equation}

In general, the integral in equation \ref{i} is calculated
numerically, but for $\delta\ll 1$ one can decompose
$[1-\delta\cos\lambda]^{-3/2} \approx 1 + 3/2 \delta\cos\lambda$ and
obtain $I(\delta, \beta) \approx (3/8) \delta \sin 2\beta$.  When this
approximation holds, that is when $R \gg R_1$ or $R\ll R_1$, the
precession frequency for the second ring is
\begin{equation}
\frac{\omega_p}{\Omega_K} \approx -
\frac{3 M_1}{4 M_{\rm BH}} \;\cos \beta\; \frac{R^3
R_1^2} {\left[R^2 + R_1^2\right]^{5/2}}\;.
\label{omegapa}
\end{equation}

Few estimates can now be made. First of all, $\omega_p$ reaches a
maximum at radius $R_m$ at which $R_m/R_1 = \sqrt{3/7}$. The maximum
growth rate of the warp is thus
\begin{equation}
\hbox{max}\;|\omega_p| \simeq 0.085 \Omega_K \; \cos
\beta\frac{M_1}{M_{\rm BH}}\;.
\label{maxomk}
\end{equation}
$M_1/M_{\rm BH}$ may be expected on the order of a percent or so since
this is when the accretion disks become self-gravitating
\citep[e.g.,][]{NC04}, and the resulting stellar mass would probably
be of that order too. In this case one notices that to produce a
sufficiently large warp one has to wait for at least $\sim 10 \times
M_{\rm BH}/M_1 \sim 1000 \Omega_K^{-1}$ or $\sim 150$ orbital times at
the radius of the maximum warp $R_m$. While this is a fairly long
time, it is still much shorter than the corresponding disk viscous
time (cf., e.g., Fig. 2 in \cite{NC04}).

The assimptotic dependence of precession frequency on radius $R$ is
\begin{eqnarray}
\label{ass1}
\omega_p \rightarrow \; - \frac{3 M_1}{4 M_{\rm BH}} \cos
\beta\;\frac{R^3}{R_1^3}
\; \Omega_K 
\quad \hbox{for}\; R \ll R_1 \;, \quad \hbox{and}
\\ 
\omega_p
\rightarrow \; - \frac{3 M_1}{4 M_{\rm BH}} \cos
\beta \; \frac{R_1^2}{R^2} \;
\Omega_K  \quad
\hbox{for}\; R \gg R_1 \;.
\label{ass}
\end{eqnarray}
A thin massive stellar ring would thus only warp a range of radii,
leaving the portions of the disk much internal and also much external
to it unaffected.

\subsection{Test disk warping}\label{sec:testdisk}

We now want to calculate the shape of a light non self-gravitating
disk warped by a massive ring $R_1$. The disk is treated as
a collection of rings with different radii $R$ and negligible
mass. We assume that the mass of the whole disk, $M$, is negligible
in comparison with the mass of the ring, $M \ll M_1$, and that
therefore the massive ring's orientation (angular momentum) does not
change with time.

In application to the real disks, it should be remembered that any
attempt to warp a disk will be resisted by disk viscous forces
\citep[e.g.,][]{Bardeen75} that transfer the angular momentum through
the disk. However, we are interested in the outermost regions of AGN
accretion disks where they ``connect'' to the galaxy. On these, a
tenth of a parsec to 100 parsec scales (the range depends on $M_{\rm
BH}$), the usual $\alpha$-disk viscosity \citep{SS73} becomes
ineffective. The viscous time scale becomes too long and the disks are
believed to be prone to self-gravity
\citep[e.g.,][]{Paczynski78,Shlosman89,Collin99,Goodman03}. Therefore
we are justified in neglecting the restoring viscous forces. A much
more important effect will be the restoring force from the
self-gravity of the disk that is being warped, but we differ the study
of this and other non-linear effects to a future paper.

It is convenient to work with angle $\beta$, already introduced, and
one additional angle, $\gamma$. Recall that angle $\beta$ is the local
angle of the ring's tilt to the $z$-axis and it remains constant in
the test particle regime. Angle $\gamma$ is needed to introduce the
projections of the unit tilt vector normal to the ring, $l(R, t)$, on
the $x$ and $y$ axes \citep{Pringle96}:
\begin{equation}
\vec{l} = \; \left(\cos\gamma \sin\beta, \sin\gamma \sin\beta,
\cos\beta\right)\;.
\label{defl}
\end{equation}
Angle $\gamma$ therefore describes the precession of each ring around
the $z$-axis. With the chosen coordinate system, at time $t=0$,
$\gamma = \pi/2$ (and we get, in accord with conventions of \S
\ref{sec:rings}, $L_y(0) = L \sin\beta$, $L_x(0)=0$). As each of the
disk rings precesses,
\begin{equation}
\gamma = \pi/2 + \omega_p t\;.
\end{equation}

It is useful to write the expression for the unit tilt vector in the
$(x', y', z')$ coordinate system rigidly bound to the initial disk
plane:
\begin{eqnarray}
l'_x = \sin\beta \cos\gamma \;,\label{deflpx}\\
l'_y= -\cos\beta\sin\beta(1-\sin\gamma) \;, \\
l'_z = \cos^2\beta +\sin^2\beta\sin\gamma\;.
\label{deflpz}
\end{eqnarray}
Note that if $\beta=0$, $l'_x = l'_y =0$, e.g. the rings are never
tilted with respect to the $z=z'$ axis, as it should. Also, when
$\gamma = \pi/2$, $\vec{l'}$ indeed coincides with the $z'$-axis,
i.e. the disk is flat.

Using equations \ref{deflpx}-\ref{deflpz}, we can now calculate the
shape of the warped disk in that system given the function
$\gamma(R,t)$. To accomplish this, one first introduces the azimuthal
angle $\phi$ on the surface of the ring. The coordinates of the points
on the ring, $\vec{r}$, are then given by equations 2.2 and 2.3 in
\cite{Pringle96}. The corresponding coordinates in the primed system
of reference are easily obtained by $x' = (\vec{r} \vec{e_x}')$, etc.,
where $\vec{e_x}'$ and so on are the unit coordinate vectors of the
primed system. The result is:
\begin{eqnarray}
\frac{x'}{R} = \cos\beta \cos\gamma\sin\phi + \sin\gamma\cos\phi\;,\\
\frac{y'}{R} = -\cos\beta\cos\gamma\cos\phi +
\sin\phi(\cos^2\beta\sin\gamma + \sin^2\beta)\;,\\ \frac{z'}{R} =
-\sin\beta\left[\cos\gamma\cos\phi +
(1-\sin\gamma)\cos\beta\sin\phi\right]\;.
\end{eqnarray}

\subsection{An example}\label{sec:example}

\begin{figure*}
\centerline{\psfig{file=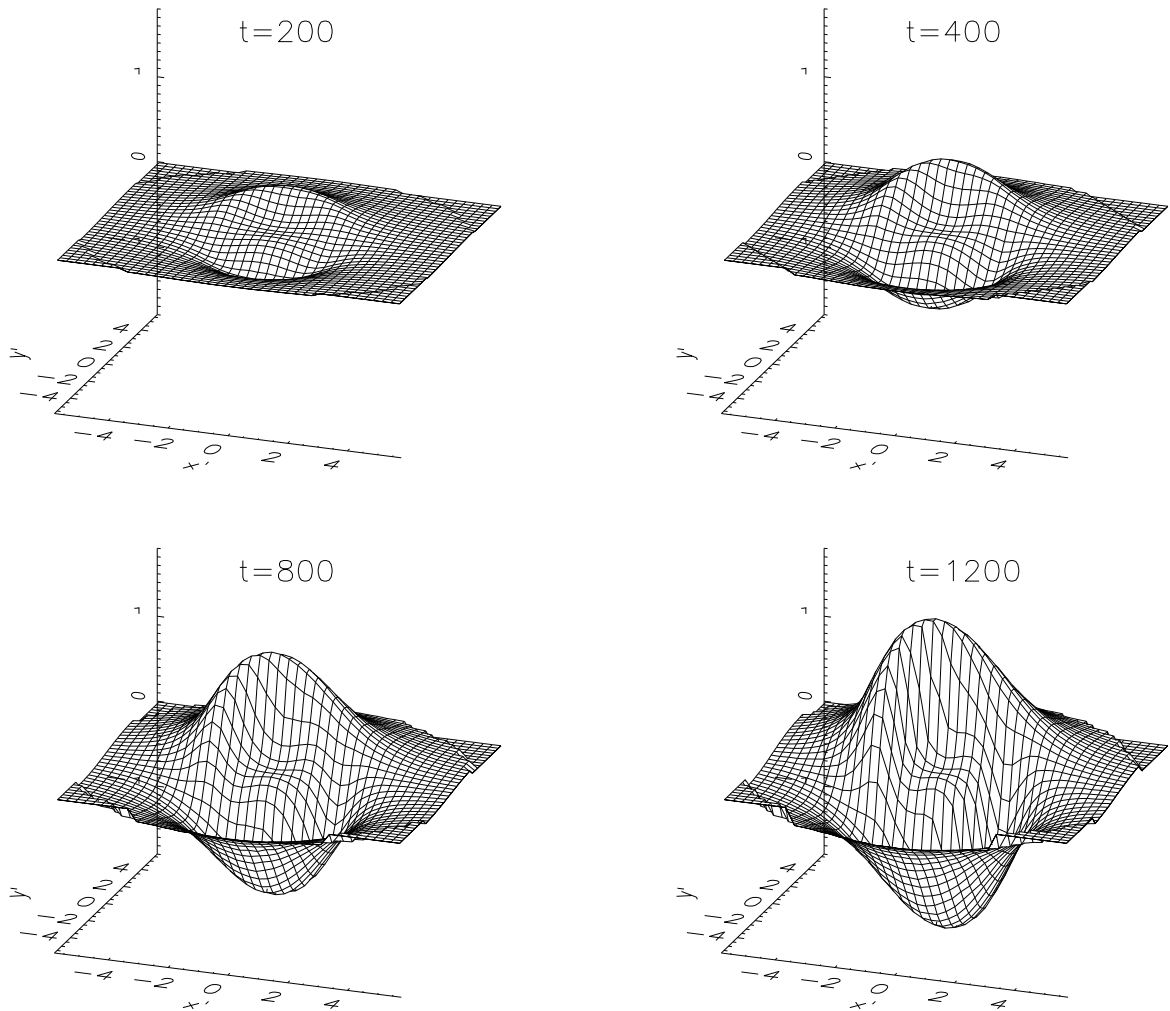,width=1\textwidth,angle=0}}
\caption{Snapshots of the shape of a massless accretion disk warped by
a stellar ring of radius $R=2$ inclined at angle $\beta=\pi/4$ with
respect to the disk. The snapshots are for four different time as
indicated on the top of each panel. The time unit is $1/\Omega_K(R=1)
= 1400 \; r_{pc}^{3/2} M_8^{-1/2}$ years. At times larger than those
used in the Figure, the disk becomes warped so much that, looking from
its initially non-warped plane, the plane equation $z'(x',y')$ becomes
a multiple-valued function for some $(x',y')$. In reality non-linear
effects will limit the growth of the warp.}
\label{fig:fig1}
\end{figure*}

To illustrate the results, we calculate the precession rates
$\omega_p(R)$ for the following case, $R_1 = 2$, $M_1 = 0.01 M_{\rm
BH}$, $\beta = \pi/4$. The resulting shape of the disk in the original
un-warped disk plane is plotted in Figure \ref{fig:fig1}. The time is
in units of $\Omega_K^{-1}$ at $R=1$, i.e. $t=1$ corresponds to time
$t=1400 \; r_{pc}^{3/2} M_8^{-1/2}$ years, where $r_{pc}$ is the
distance in units of parsec and $M_8 = M_{\rm BH}/10^8 \msun$.  Note
that the warp is strongest around radius $R\sim R_1=2$, as
expected. The inner disk is hardly tilted, which is not surprising
given that $\omega_p(R) \rightarrow 0$ for small $R$ (equation
\ref{ass1}). Same is true for the outer radii, where the small tilt of
the original plane can be seen on the edges of the disk (we do not
calculate the tilt beyond $R=5.2$ and hence the original disk plane
is still seen on the edges of the Figure).

\begin{figure}
\centerline{\psfig{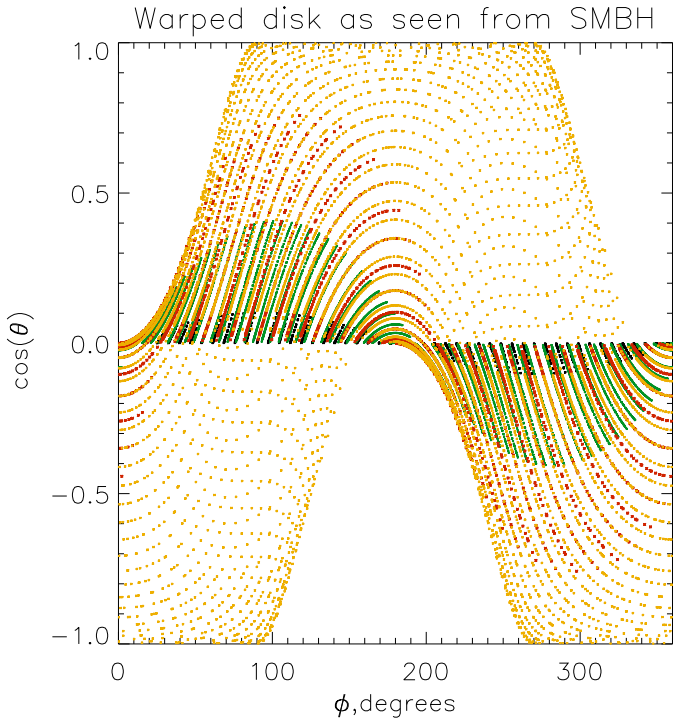}}
\caption{The accretion disk surface as seen from the SMBH
location for four different times: $t=100, 400, 800, 3200$ for the
black, green, red and yellow dots, respectively. Note that at the
largest time most of the available solid angle is obscured by the
strongly warped disk.}
\label{fig:fig2}
\end{figure}

\section{Discussion}

\subsection{Obscuration of the central engine in AGN}

We believe that the gravitational disk warping due to non-spherical
mass distribution within the SMBH sphere of influence is a common
occurrence in real AGN. It is hard to see why AGN disks should always
form in one plane and even why they should be planar when they are
born \citep{Phinney89}. The time scales for development of strong
warps are short or comparable to characteristic AGN lifetimes (which
are thought to be in the range of $10^7-10^8$ years). We thus expect
that the outer edges of the disk will obscure a significant fraction
of the sky as seen from the central source and hence be directly
relevant to the unification schemes of AGN \citep{Antonucci93}.

Figure \ref{fig:fig2} shows the warped disk surface from Figure
\ref{fig:fig1} at different times as seen from the origin of
coordinates. The vertical axis shows $\cos\theta \equiv z'/\sqrt{
(x')^2 + (y')^2 + (z')^2}$ and the horizontal one shows $\phi$ defined
previously. The shape of the disk at different times is shown with
different colors. At time $t=0$ the disk is flat and its projection is
simply $\cos\theta=0$ for all $\phi$. At later times the rings of the
disk near $R = R_1 = 2$ precess and bulge out of the initial disk
plane. With time the fraction of the sky obscured from the central
engine becomes greater than a half. In reality non-linear effects and
interaction of the disk with AGN winds and radiation, etc., will
become important in shaping the disk for significant warps.

\subsection{The non-linear evolution}

In general, the non-linear stage of the evolution of the system of
disks or rings of stars and molecular gas is far from a thin disk as
long as collisions are of minor importance. We have already explored
the non-linear evolution of collisionless systems with N-body codes
and the results will be reported elsewhere. When two massive disks
warp each other, mixing occurs when $\gamma-\pi/2$ becomes large, and
the resulting configuration reminds that of a torus. 

Now, inelastic dissipative collisions between gas clumps will
eventually become important. Relaxation of a collisionless N-body
system leads to non-circular orbits, and hence different disk rings
will start to overlap. Collisions should then tend to destroy the
random motions and should establish a common disk ``plane''. However,
such a disk will normally be warped itself.  Therefore inelastic
dissipative collisions do not necessarily turn off the obscuration in
our model. Further, if molecular gas clumps are constantly supplied
from the outside and come with fluctuating angular momentum, the disk
may never arrive in a flat thin configuration \citep[see also
][]{Phinney89}.

\cite{Nenkova02} and \cite{Risaliti02} convincingly argued that the
AGN obscuring medium must not be uniform in density. This does not
contradict our model at all since AGN accretion disks on large
(e.g. 0.1 pc and beyond) scales are self-gravitating unless the
accretion rates are tiny \citep[e.g., ][]{Shlosman89}. 

The source of warping potential does not have to be a
thin stellar ring or a disk: it may be any non-spherical distribution
of stars in the SMBH vicinity that retains a non-zero quadrupole
moment; it can also be a second (smaller) super-massive black hole
during a merger of two galaxies.

Both the collision-dominated \citep{Krolik88} and the stellar-feedback
inflated torii \citep{Wada02} share a common starting point: the torus
is the result of some internal disk physics. If our interpretation
applies, AGN torii lend their existence to the way in which the cold
gas arrives in the central part of the galaxy.  Compared with the
model of \cite{Krolik88}, large random speeds for the cold gas are not
required in our model. Although the gas may be quite high up away from
the original plane of the disk, the disk is locally coherent and
thin. High speed elastic collisions between molecular clouds are thus
not needed to explain obscuration of AGN in our model.

There are also two other mechanisms potentially able to produce warped
disks at parsec and beyond scales around AGN. As mentioned in the
X-ray binaries context, accretion disks develop twists and warps due
to instabilities driven by X-ray heated wind off the disk surface
\citep{SM94}. Same can be achieved by the radiation pressure force
from the central source \citep{Pringle96}. However, it seems that the
majority of disks in X-ray binaries are either not warped or warped
not strongly enough \citep{Ogilvie01} to provide the large obscuration
needed for the AGN unification schemes. In contrast, the warping
mechanism discussed here is not applicable to X-ray binary systems,
and hence it may be natural that AGN disks are stronger warped than
X-ray binary disks on appropriately scaled distances from the center.

\del{However in both cases the
efficiency of the instability depends on the detail of the radiation
transfer from the central source to the outer disk, and have not been
explored in the context of clumpy accretion disks. Such disks will be
effectively optically thin at large distances.  In contrast, the
gravitational warping discussed here works for both gaseous continuous
disks or disks composed of molecular clumps. Finally, the wind and
radiation pressure driven warping instabilities also work for disks in
X-ray binaries. Whereas there apparently are warped disk systems, the
majority of disks are either not warped or warped not strongly enough
\citep{Ogilvie01} to provide the large obscuration needed for the AGN
unification schemes.}


\subsection{The stellar disks in \sgra}

The two young stellar disks discovered recently in \sgra\ present a
challenge to the usual star formation modes because the gas densities
required to avoid tidal shearing are many orders of magnitude larger
than the highest densities observed anywhere in the galactic molecular
clouds. On the other hand, star formation inside a massive accretion
disk is a long expected outcome of the self-gravitational instability
of such disks
\citep[e.g.][]{Paczynski78,Shlosman89,Collin99,Goodman03}.  As such,
the young stars in the \sgra\ star cluster therefore are a first
example of star formation in this extreme environment. It is very
likely that the star formation efficiency and the initial mass
function (IMF) are quite different in the immediate AGN vicinity and
elsewhere in a galaxy. The ``astro-archeology'' of \sgra\ can be used
to study these issues.

The gravitational warping effect discussed in this paper constrains
the time-averaged mass of the outer ring, $M_{\rm outer}$, since the
moment of its creation, assumed to be $t = 2 \times 10^6$ years. The
inner stellar ring is rather well defined \citep{Genzel03}, and we
thus estimate $\omega_p t$ for the inner ring to be smaller than
$\pi/4$.  Taking the radius of the inner stellar ring to be $R =
3''\simeq 4\times 10^4 R_S$ for the GC black hole, and for the outer,
$R_1 = 5''$, and using equation \ref{omegapa} with $\cos\beta=1/4$,
one obtains $M_{\rm outer}< 10^5 \msun$. Preliminary numerical N-body
simulations show this limit may be even smaller.

\subsection{Other implications for AGN}

There are clearly other observational implications of gravitationally
warped disks. For example, Narrow \fe lines, observed in many Seyfert
galaxies, can be explained with X-ray reflection off such warped cold
disks. In addition, warped disks will yield different coherent paths
for maser amplification than flat disks do, which may be a part of the
explanation for the complexity of the observed AGN maser emission.

\section{conclusions}

We argued that clumpy self-gravitating accretion disks in AGN are
generically strongly warped. We believe such warping should be an
integral part of the explanation for the AGN unification schemes. Our
model provides arguably the easiest way to obscure the central engine
without the need to lift cold gas clouds off the disk plane high up
via elastic collisions, supernova explosions or winds.

\section{ACKNOWLEDGMENTS}

The author thanks Jorge Cuadra and Walter Dehnen for their help with
numerical simulations that motivated this semi-analytical paper, and
for very useful discussions. In addition, the author benefited from
discussions with Andrew King and Friedrich Meyer.

\bibstyle{mn2e}

\end{document}